\documentclass[twocolumn,showpacs,groupedaddress]{revtex4}
\usepackage{graphicx}

\begin{document}

\draft
\title{Entanglement distillation using particle statistics}
\author{X. L. Huang, L. H. Cheng, and X. X. Yi}
\affiliation{Department of physics, Dalian University of
Technology, Dalian 116024, China}

\date{\today}

\begin{abstract}
We extend the idea of entanglement concentration  for pure
states(Phys. Rev. Lett. {\bf 88}, 187903) to the case of mixed
states. The scheme works only with particle statistics and local
operations, without the need of any other interactions. We show
that the  maximally entangled state can be distilled out when the
initial state is pure, otherwise the entanglement of the final
state is less than one. The distillation efficiency is a product
of the diagonal elements of the initial state, it takes the
maximum  $50\%$, the same as the case for pure states.
\end{abstract}

 \maketitle

Entanglement shared between two distinguishable particles is
generally an advantage for quantum information processing, while
indistinguishability prevents us from addressing the particles
separately that seems to be a disadvantage in information
processing. It is only recently that researchers have started to
investigate the use of particle statistics (both bosonic and
fermionic) for quantum information
processing\cite{omar02,paunkovic02,bose03}. It was shown that
quantum statistics can lead to an effective interaction between
internal and external degrees of freedom of
particles\cite{omar02}, and then result in space-spin entanglement
transfer. The other useful tasks such as entanglement
concentration\cite{paunkovic02} and state
discrimination\cite{bose03}could also be accomplished, even if
nonoptimally, using only the effects of quantum statistics,
without the need of any other interactions. These proposals purely
based on particle statistics differ significantly from some
previous suggestions, where particle statistics together with
interparticle interactions \cite{loss98,costa01,yu03} were used
for quantum information processing.

The entanglement distillation/concentration is essential in
quantum communication and computation, it increases the
entanglement shared between distant parties, and consequently
could better the performances in quantum information processing.
As entanglement can not be increased by local operations and
classical communication\cite{plenio98}, the
distillation/concentration operation is to distill/concentrate
entanglement locally from a large to a smaller number of
pairs\cite{gisin96, bennett961,bennett962,deutsch96, bose99,
hardy00, horodecki97, thew01, kwiat01, pan01}.

\begin{figure}
\includegraphics*[width=0.8\columnwidth,
height=0.8\columnwidth]{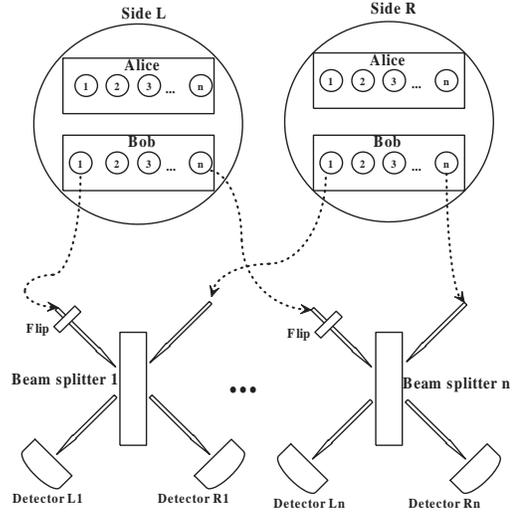} \caption{ An schematic
illustration of the scheme. Alice and Bob start with two pairs of
entangled systems at side $L$ and $R$, respectively. At each side,
Alice and Bob share $n$ pairs of particles. After a set of local
operations including spin flip and 50/50 beam splitter, path
measurements, and classical communications, states with more
entanglement can be distilled out from the initial state. The
scheme works with particle statistics solely, without the need of
other interactions.} \label{fig1}
\end{figure}

Quantum information processing based on particle statistics is
very useful for tasks implemented with identical particles. In
this paper, we extend the protocol proposed in
Ref.\cite{paunkovic02} for pure states to mixed states. To start
with, we recall and summarize the entanglement concentration
protocol using only particle statistics as follows. The initial
state is prepared at two sides labelled by $L$ and $R$. In each
side, two parties Alice (A) and Bob (B) share $n$ pairs of
particles. The protocol includes the following four steps. (1)
Bring each of Bob's particles (from both sides of $L$ and $R$)
into a 50/50 beam splitter; (2) Before the particles from the side
$L$ fall into the beam splitter, a specific unitary transformation
that flips the particle's spin is applied to each of the
particles; (3) Make a specific measurement on the output
particles, and discard those which bunch(for
fermions)/antibunch(for bosons); (4) Repeat the above three steps
for all particles with Bob. This was schematically illustrated in
figure \ref{fig1}.
We now apply the above steps to mixed states for fermions, the
results can be straightforwardly generalized to the case of
bosons. As the authors did in Ref.\cite{paunkovic02}, we consider
the entanglement in the internal degrees of freedom of the
particles, for example the spin in the case of fermions. In either
sides $L$ or $R$, Alice and Bob share $n$ pairs of particles, the
initial state at side $I$ ($I=R,L$), distributed between Alice and
Bob, in basis of $\{|A\tilde{\uparrow}\rangle^n_I
|B\tilde{\downarrow}\rangle^n_I, |A\tilde{\downarrow}\rangle^n_I
|B\tilde{\uparrow}\rangle^n_I \}$ is
\begin{equation}
\rho^n_I=\left(
               \matrix{
               a & c   \cr
               c^{*} & b
             }
        \right). \label{inistates1}
\end{equation}
where $|A\tilde{\uparrow}\rangle_L^n=|\underbrace{\uparrow
\uparrow ...\uparrow}_n\rangle_{L,A}$ denotes $n$ spin-up
particles at $L$ side with Alice. The other notations in the basis
are similar to $|A\tilde{\uparrow}\rangle_L^n$.
 The total initial state under consideration is then
\begin{eqnarray}
  \rho^n=\rho_L^n\otimes\rho_R^n=\left(
               \matrix{
               a^2 & ac & ac & c^2 \cr
               ac^* & ab & |c|^2 & bc\cr
               ac^* & |c|^2 & ab  & bc\cr
               c^{*2} & bc^* & bc^* & b^2
             }
        \right). \label{inistate1}
\end{eqnarray}
Here, the basis chosen is
\begin{eqnarray}
|\alpha\rangle&\equiv&|A\tilde{\uparrow}\rangle_L^n|A\tilde{\uparrow}
\rangle_R^n|B\tilde{\downarrow}\rangle_L^n|B\tilde{\downarrow}\rangle_R^n,\nonumber\\
|\beta\rangle&\equiv&|A\tilde{\uparrow}\rangle_L^n|A\tilde{\downarrow}
\rangle_R^n|B\tilde{\downarrow}\rangle_L^n|B\tilde{\uparrow}\rangle_R^n,\nonumber\\
|\gamma\rangle&\equiv&|A\tilde{\downarrow}\rangle_L^n|A\tilde{\uparrow}
\rangle_R^n|B\tilde{\uparrow}\rangle_L^n|B\tilde{\downarrow}\rangle_R^n,\nonumber\\
|\kappa\rangle&\equiv&|A\tilde{\downarrow}\rangle_L^n|A\tilde{\downarrow}
\rangle_R^n|B\tilde{\uparrow}\rangle_L^n|B\tilde{\uparrow}\rangle_R^n.
\end{eqnarray}
With this total initial state, next step we let the first pair of
particles(labelled by 1 in the circle) with Bob go to the beam
splitter 1, after having the specific unitary transformation in
step (2), the total state still takes the form of
Eq.(\ref{inistate1}), but was written in basis of $
 \{ |A\tilde{\uparrow}\rangle_L^n|A\tilde{\uparrow}
\rangle_R^n|B\tilde{\uparrow}\rangle_L^n|B\tilde{\downarrow}\rangle_R^n,$
$|A\tilde{\uparrow}\rangle_L^n|A\tilde{\downarrow}
\rangle_R^n|B\tilde{\uparrow}\rangle_L^n|B\tilde{\uparrow}\rangle_R^n,$
$ |A\tilde{\downarrow}\rangle_L^n|A\tilde{\uparrow}
\rangle_R^n|B\tilde{\downarrow}\rangle_L^n|B\tilde{\downarrow}\rangle_R^n,$
$ |A\tilde{\downarrow}\rangle_L^n|A\tilde{\downarrow}
\rangle_R^n|B\tilde{\downarrow}\rangle_L^n|B\tilde{\uparrow}\rangle_R^n.
\}$ Similarly, the state of particles after the first pair having
passed through the 50/50 beam splitter can be written in the same
form as in Eq.(\ref{inistate1}) in the Hilbert space spanned by
\begin{eqnarray}
|\alpha_p\rangle&\equiv&|A\tilde{\uparrow}\rangle_L^n|A\tilde{\uparrow}
\rangle_R^n|B_1\rangle_{L1R1}|B\tilde{\uparrow}\rangle_L^{n-1}|B\tilde{\downarrow}\rangle_R^{n-1},\nonumber\\
|\beta_p\rangle&\equiv&|A\tilde{\uparrow}\rangle_L^n|A\tilde{\downarrow}
\rangle_R^n|B\tilde{\uparrow}\rangle_L^n|B\tilde{\uparrow}\rangle_R^n,\nonumber\\
|\gamma_p\rangle&\equiv&|A\tilde{\downarrow}\rangle_L^n|A\tilde{\uparrow}
\rangle_R^n|B\tilde{\downarrow}\rangle_L^n|B\tilde{\downarrow}\rangle_R^n,\nonumber\\
|\kappa_p\rangle&\equiv&|A\tilde{\downarrow}\rangle_L^n|A\tilde{\downarrow}
\rangle_R^n|B_2\rangle_{L1R1}|B\tilde{\downarrow}\rangle_L^{n-1}|B\tilde{\uparrow}\rangle_R^{n-1},
\label{basis1}
\end{eqnarray}
where $|B_1\rangle_{L1R1}=\frac 1 2
i(|B\uparrow\rangle_{L1}|B\downarrow\rangle_{L1}+|B\uparrow\rangle_{R1}|B\downarrow\rangle_{R1})
+ \frac 1
2(|B\uparrow\rangle_{L1}|B\downarrow\rangle_{R1}+|B\downarrow\rangle_{L1}|B\uparrow\rangle_{R1}),$
and $|B_2\rangle_{L1R1}=-\frac 1 2
i(|B\uparrow\rangle_{L1}|B\downarrow\rangle_{L1}+|B\uparrow\rangle_{R1}|B\downarrow\rangle_{R1})
+ \frac 1
2(|B\downarrow\rangle_{L1}|B\uparrow\rangle_{R1}+|B\uparrow\rangle_{L1}|B\downarrow\rangle_{R1}).$
When  the particle had passed  through the beam splitter, Bob
performs a path measurement on the first pair with assumption that
the detectors do not absorb the particles and do not disturb their
internal(spin) degrees of freedom. Discarding those particles
which bunch, we arrive at
\begin{eqnarray}
  \rho^n_1=N_1^2\left(
               \matrix{
               \frac 1 2 a^2 & \frac {1} {\sqrt 2}ac & \frac {1} {\sqrt 2}ac & \frac 1 2c^2 \cr
               \frac {1} {\sqrt 2}ac^* & ab & |c|^2 &\frac {1} {\sqrt 2} bc\cr
               \frac {1} {\sqrt 2}ac^* & |c|^2 & ab  & \frac {1} {\sqrt 2}bc\cr
               \frac 1 2c^{*2} &\frac {1} {\sqrt 2} bc^* & \frac {1} {\sqrt 2}bc^* & \frac 1 2b^2
             }
        \right), \label{finstate1}
\end{eqnarray}
where $N_1=(\frac 1 2 a^2+\frac 1 2 b^2+2ab)^{-1/2}$, and the
corresponding basis is the same as that in Eq.(\ref{basis1}), but
the bunching terms in $|B_1\rangle_{L1R1}$ and
$|B_2\rangle_{L1R1}$ were discarded. The probability of having
state Eq.(\ref{finstate1}) is $(0.5+ab)$, it arrives at the
maximum $3/4$ with $a=b=1/2$. Eq.(\ref{basis1}) tells us that the
bunching is only related to $|\alpha_p\rangle$ and
$|\kappa_p\rangle$, and these two terms have probability $1/2$ of
antibunching. Thus elements of the density matrix
Eq.(\ref{finstate1}) differ from the initial state upon the
normalization factor $1/\sqrt{2}$ or $1/2$, depending on the total
probability of antibunching. With the antibunching state as an
initial state, next we let the second pair of Bob's
particles(labelled by 2 in the circle) pass through another 50/50
beam splitter and perform the same measurement, keeping again the
antibunching results. Having completed the step (4) and absorbed
all these $(n-1)$ particles, we get the final state
\begin{eqnarray}
  \rho^n_n=N_n^2\left(
               \matrix{
               (\frac 1 2)^n a^2 & (\frac {1} {\sqrt 2})^n ac & (\frac {1} {\sqrt 2})^n ac & (\frac 1 2)^nc^2 \cr
               (\frac {1} {\sqrt 2})^n ac^* & ab & |c|^2 &(\frac {1} {\sqrt 2})^n bc\cr
               (\frac {1} {\sqrt 2})^nac^* & |c|^2 & ab  & (\frac {1} {\sqrt 2})^nbc\cr
               (\frac 1 2)^n c^{*2} &(\frac {1} {\sqrt 2})^n bc^* & (\frac {1} {\sqrt 2})^n bc^* & (\frac 1 2)^nb^2
             }
        \right)\label{finstate2}
\end{eqnarray}
with $N_n=((\frac 1 2)^n a^2+(\frac 1 2)^n b^2+2ab)^{-1/2}$, and
the basis
\begin{eqnarray}
|\alpha_f\rangle&\equiv&|A\tilde{\uparrow}\rangle_L^n|A\tilde{\uparrow}
\rangle_R^n|B_{triplet}\rangle,\nonumber\\
|\beta_f\rangle&\equiv&|A\tilde{\uparrow}\rangle_L^n|A\tilde{\downarrow}
\rangle_R^n|B\tilde{\uparrow}\rangle_L|B\tilde{\uparrow}\rangle_R,\nonumber\\
|\gamma_f\rangle&\equiv&|A\tilde{\downarrow}\rangle_L^n|A\tilde{\uparrow}
\rangle_R^n|B\tilde{\downarrow}\rangle_L|B\tilde{\downarrow}\rangle_R,\nonumber\\
|\kappa_f\rangle&\equiv&|A\tilde{\downarrow}\rangle_L^n|A\tilde{\downarrow}
\rangle_R^n|B_{triplet}\rangle.
\end{eqnarray}
Here, $|B_{triplet}\rangle=\frac
{1}{\sqrt{2}}(|B\uparrow\rangle_{L1}|B\downarrow\rangle_{R1}+|B\downarrow\rangle_{L1}|B\uparrow\rangle_{R1})$
The probability of having this state is $p_f=((\frac 1 2)^n
a^2+(\frac 1 2)^n b^2+2ab).$ With $n\rightarrow\infty$, the
probability $p_f=2ab$, while the entanglement measured by Wootters
concurrence is $|c|^2/ab$. The entanglement of the final state
reaches its maximal value 1 with $|c|^2=ab$, i.e., in the pure
state case. For mixed states, because $|c|^2<ab$ required by
Eq.(\ref{inistates1}), the entanglement of the final state is
always less than unity, and it is of relevance to the coherence of
state $\rho_I^n$ measured by $|c|$ in Eq.(\ref{inistates1}). On
the other hand, the state Eq.(\ref{inistates1}) itself is an
entangled state, the entanglement measured by Wootters concurrence
is $2|c|$. To get more entanglement in the final state, it is
required that $|c|>2ab$. Entangled state with $|c|\leq 2ab$ could
not be distilled, this is similar to the non-distillable
entanglement called bound entanglement.

In conclusion, we have extended the entanglement concentration
protocol to the case of mixed states, which uses only the effects
of particle statistics. The maximal entanglement after
distillation depends on the off-diagonal elements of the initial
state that usually is a measure of coherence. The distillation
efficiency only depends on the diagonal element of the initial
state, it takes $50 \%$ for finite number of particles and tends
to $25\%$ for infinite large states.

\ \ \\
This work was supported
by NCET of M.O.E, and NSF of China Project No. 10305002.\\


\begin{references}

\bibitem{omar02} Y. Omar, N. Paunkovic, S. Bose, and V. Vedral,
Phys. Rev. A {\bf 65}, 062305(2002).

\bibitem{paunkovic02} N. Paunkovic, Y. Omar, S. Bose, and V.
Vedral, Phys. Rev. Lett. {\bf 88}, 187903(2002).

\bibitem{bose03} S. Bose, A. Ekert, Y. Omar, N. Paunkovic, and V.
Vedral, Phys. Rev. A {\bf 68}, 052309(2003).

\bibitem{loss98} D. Loss and D. P. DiVincenzo, Phys. Rev. A {\bf
57}, 120(1998).

\bibitem{costa01} A. T. Costa, Jr. and S. Bose, Phys. Rev. Lett.
{\bf 87}, 277901(2001).

\bibitem{yu03} A. Yu. Kitaev, Ann. Phys. (N.Y.){\bf 303}, 2(2003).

\bibitem{plenio98} M. Plenio and V. Vedral, Contemp. Phys. {\bf
39}, 431(1998).

\bibitem{gisin96} N. Gisin, Phys. Lett. A {\bf 210}, 151(1996).

\bibitem{bennett961} C. H. Bennett, H. J. Bernstein, S. Popescu.
and B. Schumacher, Phys. Rev. A {\bf 53}, 2046(1996).

\bibitem{bennett962} C. H. Bennett, D. P. Divincenzo, J. A.
Smolin, and W. K. Wootters, Phys. Rev. A {\bf 54}, 3824(1996).

\bibitem{deutsch96} D. Deutsch {\it etal.,} Phys. Rev. Lett. {\bf
77}, 2818(1996).

\bibitem{bose99} S. Bose, V. Vedral, and P. L. Knight, Phys. Rev.
A {\bf 60}, 194{1999}.

\bibitem{hardy00} L. Hardy and D. D. Song, Phys. Rev. A {\bf 62},
052315(2000).

\bibitem{horodecki97}M. Horodecki, P. Horodecki, and R. Horodecki,
Phys. Rev. Lett. {\bf 78}, 574(1997).

\bibitem{thew01} R. T. Thew and W. J. Munro, Phys. Rev. A {\bf
63}, 030302(2001).

\bibitem{kwiat01} P. G. Kwiat, S. Barraza-Lopez, A. Stefanov, and
N. Gisin, Nature(London) {\bf 409}, 1014(2001).

\bibitem{pan01} J. W. Pan, C. Simon, C. Brukner, and A. Zeilinger,
Nature(London) {\bf 410}, 1067(2001).

\end{references}
\end{document}